\documentclass[pra,twocolumn,floats,showpacs,preprintnumbers,tighten,epsfig,superscriptaddress]{revtex4}

\usepackage{graphicx}
\usepackage{dcolumn}
\usepackage{bm,color}
\usepackage{amsmath}
\usepackage{amsfonts}
\usepackage{amssymb}

\usepackage{ulem} 

\newcommand{\red}{\color{red}}

\newcommand{\blk}{\color{black}}
\newcommand{\blu}{\color{blue}}
\definecolor{ngreen}{rgb}{0.3,0.7,0.3}

\definecolor{golden}{rgb}{0.8,0.6,0.1}

\renewcommand\sout[1]{}

\renewcommand\blu{\blk}
\renewcommand\red{\blk}

\begin{document}

\title{Nonlocal signaling in the configuration space model of quantum-classical interactions}

\author{Michael J.  W. Hall}
\affiliation{Centre for Quantum Dynamics, Griffith University, Brisbane QLD 4111, Australia}
\author{Marcel Reginatto}
\affiliation{
Physikalisch-Technische Bundesanstalt, 38116 Braunschweig, Germany}
\author{C.~M.~Savage}
\affiliation{
Department of Quantum Science, Research School of Physics and Engineering, Australian National University, Canberra ACT 0200, Australia}

\begin{abstract}
When interactions are turned off, the theory of interacting quantum and classical ensembles due to Hall and Reginatto is shown to suffer from a nonlocal signaling effect that is effectively action at a distance. This limits the possible applicability of the theory. In its present form, it is restricted to those situations in which interactions are always on, such as classical gravity interacting with quantized matter.
\end{abstract}

\maketitle

Hall and Reginatto have developed a theory that attempts to unify quantum and classical systems, \blu and interactions between them, \blk within a single theoretical structure \cite{Hall and Reginatto PRA, Hall PRA, Reginatto and Hall JP}. It has been applied to simple systems, such as a quantum and a classical particle interacting via a harmonic potential, to various idealized measurement scenarios \cite{Hall and Reginatto PRA, Hall PRA,Chua}, and to classical spacetimes gravitationally interacting with quantum matter fields \cite{Hall and Reginatto PRA, Albers}.

In its most general form, the theory allows \blu action at a distance, \blk in which changes in one system induce changes in  another system with which it previously interacted, even after the interaction is turned off. To address this, the original theory assumed that only classical position variables were \blu {\it directly} \blk observable \blu \cite{Hall PRA}. \blk This \blu assumption is sufficient to prevent \blk a change in one system being instantaneously observable in another. \blu Thus, for example, it rules out the `ghost interaction' effect noted by Salcedo \cite{Salcedo}, which requires that the classical kinetic energy be directly observable. \blk

However, here we show that \blu the above assumption \blk does not prevent changes becoming \blu statistically \blk observable over time. Specifically, the change can manifest in the second time derivative of the position variance. \blu This allows the possibility of nonlocal signaling, in which the position statistics of a classical ensemble are altered by operations carried out on a distant quantum ensemble with which it has formerly interacted.  \blk

Consequently, in its current form \blu the theory \blk is not a viable candidate for a fundamental theory of the mesoscopic quantum-classical interface. \blk \sout{as discussed by Chua et al. \cite{Chua}.} However, the version of the hybrid theory for a classical gravitational field coupled to quantized matter fields is unaffected \blu by this result, \sout{by ghost interactions \cite{Albers}.} \blk because the gravitational interaction cannot be turned off. The direct coupling of the classical metric tensor to the quantum fields implies there is no sense in which the corresponding ensemble
Hamiltonian can be reduced to a simple sum of a classical
and a quantum contribution with no interaction \blu \cite{Hall PRA}. \blk

Other types of quantum-classical hybrid theories do not allow non-local signalling, for example: the mean-field theories of Boucher and Traschen \cite{Boucher} and of Elze \cite{Elze}. However, these theories have different problems \cite{Salcedo,Boucher}.

In the interest of brevity we provide only a brief outline of the hybrid theory, \blu sufficient to demonstrate the above results. \blk  A complete explanation of the theory is presented in previous publications \cite{Hall and Reginatto PRA, Hall PRA, Reginatto and Hall JP}.

The hybrid theory \blu for interacting quantum and classical particles \blk is based on a probability density $P(q,x)$ in configuration space and on a conjugate function $S(q,x)$. The quantum particle has coordinate $q$ and mass $m_q$.  The classical particle has coordinate $x$ and mass $m_x$. \blu Evolution is generated by an \blk ensemble Hamiltonian
\begin{align} \label{Hamiltonian}
H[P,S] &= \int dq\,dx\, P\,  \left[ \frac{(\partial_q S)^2}{2m_q} + \frac{(\partial_x S)^2}{2m_x}  \right.
\nonumber  \\
& \left. + \frac{\hbar^2}{4} \frac{(\partial_q \ln P)^2}{2m_q} + V(q,x) \right] ,
\end{align}
where the potential, $V(q,x)=V_{qx}(q,x) +V_q(q) +V_x(x)$, consists of an interaction potential between the particles $V_{qx}(q,x)$, a quantum potential $V_q(q)$, and a classical potential $V_x(x)$. The observables of the theory are functionals, $A[P,S]$, \blu corresponding to  \blk ensemble expectation values. For a classical phase space function $C(x,p_x)$,  where $p_x$ is the momentum, the  expectation value is
\begin{equation} \label{cf}
\langle C \rangle = \int dq\,dx\, P \, C(x,\partial_x S)  ,
\end{equation}
while for a Hermitian operator $Q$ associated with the quantum particle, the expectation value is
\begin{equation} \label{qm}
\langle Q \rangle = \int dq\,dx\, \psi^* Q \psi ,
 \end{equation}
where $\psi = \sqrt{P} \exp(i S/\hbar)$.

We define the Poisson bracket between two functionals $A$ and $B$ as
\begin{equation} \label{poiss}
\{A,B\} = \int dq\,dx\, \left( \frac{\delta A}{\delta P} \frac{\delta B}{\delta S} -\frac{\delta B}{\delta P} \frac{\delta A}{\delta S}  \right)  ,
\end{equation}
where variational derivatives are taken. This reduces to the usual classical Poisson bracket for classical observables and to $1/(i \hbar)$ times the quantum commutator for quantum observables. An observable $B$ may generate the infinitesimal canonical transformation
\begin{equation} \label{canonical transformation}
P \rightarrow P +\epsilon \delta B / \delta S , \quad
S \rightarrow S -\epsilon \delta B / \delta P ,
\end{equation}
where $\epsilon$ is an infinitesimal number. The corresponding transformation in an observable $A$ is
\begin{equation} \label{observable transformation}
A \rightarrow A +\epsilon \{ A,B \} .
\end{equation}

The dynamical equation for a functional $A$ is \cite{Hall PRA}
\begin{equation} \label{evo}
d A/d t = \{A,H\} .
\end{equation}

The Poisson bracket between a quantum observable and a classical observable depending only on the position $x$, is zero \cite{Hall PRA}:
\begin{equation} \label{classical quantum bracket}
\{\langle C(x) \rangle, \langle Q \rangle\} = 0.
 \end{equation}
Hence a canonical transformation on the quantum sector does not change classical configuration space observables. However classical observables that are functions of the momentum may have non-zero Poisson brackets with quantum observables \blu \cite{Hall PRA,Salcedo}. Hence the hybrid theory assumes that only classical configuration space observables are \blu directly \blk measurable \cite{Hall PRA}. \blk This \blu implies via Eq.~(\ref{classical quantum bracket}) that no local transformation of the quantum system, generated by some $\langle Q\rangle$, can instantaneously change any observable expectation value $\langle C(x)\rangle$ of the classical system.  

Nevertheless one might suspect that induced changes in the classical momentum might over time appear as changes in configuration space, and hence be measurable \cite{howard}. We show that this turns out to be the case.

We will consider time derivatives of classical configuration space observables in the absence of an interaction between the classical and quantum particles. Then the ensemble Hamiltonian Eq.~(\ref{Hamiltonian}) has $V(q,x) = V_q(q) + V_c(x)$ and may be written as $H = H_q +H_c$ where $H_q$ depends only on quantum sector quantities and $H_c$ depends only on classical sector quantities. The time derivative of a classical observable $\langle C(x) \rangle$ depending on only the classical position $x$ is
\begin{equation} \label{first derivative}
\frac{d \langle C(x) \rangle}{dt} = \{ \langle C(x) \rangle , H_q +H_c \} =  \{ \langle C(x) \rangle , H_c \}  ,
\end{equation}
where we have used Eq.~(\ref{classical quantum bracket}). Hence there is no effect of the quantum sector Hamiltonian on the \blu first time derivative of the \blk classical observable $\langle C(x) \rangle$. 

Next, however, consider the second time derivative of $\langle C(x) \rangle$:
\begin{align} \label{second derivative}
\frac{d^2 \langle C(x) \rangle}{dt^2} =& \{  \{ \langle C(x) \rangle , H_q +H_c \} , H_q +H_c \}
 \nonumber \\
=& \{  \{ \langle C(x) \rangle , H_c \} , H_c \} + \{  \{ \langle C(x) \rangle , H_c \} , H_q \}
\nonumber \\
&+ \{  \{ \langle C(x) \rangle , H_q \} , H_c \} + \{  \{ \langle C(x) \rangle , H_q \} , H_q \} .
\nonumber \\
&
\end{align}
On the second line, the first term is the classical evolution, and the last two terms are zero because of Eq.~(\ref{classical quantum bracket}). The second term is not necessarily zero and is the origin of the \blu nonlocal signaling effect. \blk It may be rewritten using the Jacobi identity and Eq.~(\ref{classical quantum bracket}) as:
\begin{align} \label{second term}
\{  \{ \langle C(x) \rangle , H_c \} , H_q \}
=& - \{  \{ H_q , \langle C(x) \rangle \} , H_c \}
\nonumber \\
&- \{  \{ H_c , H_q \} , \langle C(x) \rangle \}
\nonumber \\
=& - \{  \{ H_c , H_q \} , \langle C(x) \rangle \} .
\end{align}
The Poisson bracket of the quantum and classical sector ensemble Hamiltonians is not generally \blu of the form of a quantum observable $\langle Q\rangle$, and so this term need not vanish. \blk
In that case, the second time derivative of $\langle C(x) \rangle$ may depend on parameters of the quantum Hamiltonian, $H_q$, and non-local signalling becomes possible.

We now give an explicit example for which this happens. Consider the classical position variance observable $\langle C(x) \rangle = \langle x^2 \rangle$ and the free particle ensemble Hamiltonians $H_q = \langle \hat{p}_q^2 \rangle /(2 m_q^2)$ and $H_c = \langle p_x^2 \rangle /(2 m_c^2)$, where $\hat{p}_q$ and $p_x$ are the usual quantum momentum operator and classical momentum, \blu with \blk
\begin{align} \label{momentum observables}
\langle p_x^2 \rangle =& \int dq\,dx\, P\, (\partial_x S)^2 ,
\nonumber\\
\langle \hat{p}_q^2 \rangle =& \int dq\,dx\, P\, \left[ (\partial_q S)^2 + \frac{\hbar^2}{4} (\partial_q \ln P)^2  \right] .
\end{align}

Using the first two terms of Eq.~(\ref{second derivative}),
\begin{align} \label{example}
\frac{d^2 \langle x^2 \rangle}{dt^2} =&
\{  \{ \langle x^2 \rangle , \langle p_x^2 \rangle \} , \langle p_x^2 \rangle \} / (4 m_x^2)
\nonumber \\
&+ \{  \{ \langle x^2 \rangle , \langle p_x^2 \rangle \} , \langle \hat{p}_q^2 \rangle \} / (4 m_x m_q) ,
\nonumber \\
=& 2 \langle p_x^2 \rangle / m_x^2 \nonumber \\
&+  \frac{\hbar^2}{4 m_x m_q} \int dq\, dx\, P\, x\, \partial_x ( (\partial_q \ln P )^2 ).
\end{align}
The last integral will in general give the classical quantity $d^2 \langle x^2 \rangle / dt^2$ a dependence on the mass of the quantum particle $m_q$. In principle this mass might be varied and hence change the \blu subsequent time evolution of the classical position statistics. \blk This influence in the absence of an interaction  \blu is \blk only possible when the two systems have previously interacted \blu such \blk that the configuration probability density $P(q,x)$ does not factorize. If it does factorize, $P(q,x) = P_q(q) P_x(x)$, then the integral in Eq.~(\ref{example}) is zero \blu and no influence is possible --- \blk as expected for independent systems \blu \cite{Hall PRA}. \blk

\blu Influences \blk of the kind described would be detectable in an ensemble if, \blu for example, \blk the change in the quantum system was made in only one half of the ensemble. The classical position variance $\langle x^2 \rangle$ could be measured for each sub-ensemble and a comparison made. A difference between them would show that the quantum system had been changed. \blu  More generally, nonlocal signaling would be possible via spatial separation of the classical and quantum components of such an ensemble.  An observer choosing whether or not to make a local change to the quantum members of the ensemble could then communicate a bit of information that could be determined by local measurements on the distant classical ensemble. \blk

It has been previously shown that when the potential is quadratic and the configuration probability density is Gaussian, then numerical solutions for the dynamics of the hybrid system may be found \cite{Chua}. We have used such solutions to verify that the classical position variance observable $\langle x^2 \rangle$ depends on changes in the quantum system, even after the interaction potential is turned off, so that $V_{qc}(q,x) = 0$. We interacted the systems for a certain time and then turned the interaction off. The time evolution of $\langle x^2 \rangle$ in this case was compared with that when $m_q$ was varied after the interaction was turned off. We found that the dynamics was indeed different, as expected from the previous discussion. We also found signaling when $V_q(q)$ was varied instead of $m_q$.

Nonlocal signaling may be an indication that the hybrid theory, as it stands, is {\it incomplete}. To explore this possibility, it is  useful to consider which types of systems would be assigned to the classical sector of a hybrid system. If nature provides us with classical systems, they are likely to be either massive objects (e.g., measuring devices, as in the Copenhagen interpretation of quantum mechanics) or fields that may be fundamentally classical (e.g., gravity). Nonlocal signaling does not occur in hybrid systems consisting of a classical gravitational field coupled to quantum matter, as we have already discussed. 

In the case of massive objects, a mechanism that leads to spontaneous collapse of the classical probability would suffice to exclude signaling from the theory: collapsed quantum-classical ensembles become independent and remain independent if they no longer interact \cite{Hall PRA}, in which case no signaling is possible. The spontaneous collapse of the classical probability over a sufficiently short time scale could be achieved by means of a dynamical reduction model for the classical sector. Such a model would be analogous to that introduced into quantum mechanics by Ghirardi, Rimini and Weber (GRW) with the aim of completing the quantum theory by incorporating spontaneous collapse of the wave function \cite{Ghirardi,Bassi}. While such a mechanism has not been formulated yet for the hybrid theory, it seems possible in principle.

In summary, we have reported nonlocal signaling in the Hall-Reginatto hybrid theory that prevents it from being a viable fundamental theory \blu of interacting classical and quantum particles, at least in its present form.  \blk Nevertheless, it still provides a viable framework for a fundamental theory of a classical gravitational field interacting with quantized matter \blu  
\blk \red \cite{Hall and Reginatto PRA, Albers}. \blk

M.J.W.H \blu gratefully acknowledges a \blk seminal conversation with H.M. Wiseman.

\end{document}